# Navigating the Uncharted Waters: A Gradual Approach to the Certification and Integration of Maritime Autonomous Surface Ships (MASS) in Global Maritime Operations

Hany M. Arnaoot
*Mechatronics derpartement*
*Alexandria Higher Institute of Engineering & Technology*
Alexandria, Egypt
dr.hany.arnaout@aiet.edu.eg

*Abstract*—: This paper provides a framework towards the registration and operational chic of Maritime Autonomous Surface Ships (MASS) at the global nautical stage. This framework employs a gradual, multi-stage approach beginning from the use of small autonomous vessels at controlled environments, to international waters. This will allow for the emphasis on specific issues such as control systems and sensor fusion to be established and achieved in a gradual manner. The framework incorporates human-systems to help set effective progress measures, allow for the full controlled trial to autonomous seafaring, and provide oversight at all levels of the system. The goal of this methodology is not only to encourage innovation at MASS, but also to balance the risk and safety of a newly deployed system. The emphasis on compliance and data helps refine and improve the system. With proper transition and encouragement, this approach will shift the focus from operational risk to compliance, paving the way for autonomous ships. As a result, these measures aid the heavy lifted transition of the maritime industry towards autonomous shipping while solving economic, regulatory, and technical issues.

## I. INTRODUCTION

The majority of global trade activity, approximately 90%, along with 70% of its estimated value, is undertaken through maritime transport. This mode facilitates global commerce by serving as a dominant channel of transportation for international trade, making it feasible to transport bulk and oversized items at a low price. This allows for international trade to make great strides and expand effortlessly as marine transport acts as a central artery through which goods flow through the oceans and seas. The remaining 10% of sea trade volume is further divided between transportation via railways, roads, and airways. Out of this 10%, air transport only makes up 1% of global trade volume but due to its efficiency in bolstering high value goods, it accounts for 35% of global trade value.
.

Advanced, cutting-edge technologies, including AI, have made significant advancements and it is expected that these autonomous ships will increase safety, operational efficiency, and be economically viable. With the emergence of AI-operated ships comes significant change within the maritime sector as well as the need for comprehensive legislative reform.
1898 marked the inception of drone vessels (or autonomous ships) as the first official proposal of autonomous vessel navigation came to existence. The novel patent method of and Apparatus for Controlling Mechanism of Moving Vessels or Vehicles gives detailed information regarding the automated control of vessel mechanisms, outlining the preliminary concepts of drone ships. In terms of maritime transport, this is one of the earliest known efforts to implement and develop autonomous systems.[1]

A legal and technical framework set by the IMO defines a Maritime Autonomous Surface Ship (MASS). The IMO defines it, in regards to types of ships operating with different degrees of autonomy from human control, as those ships fully controlled by human to other ships fully controlled by computer systems operated on various algorithms. By definition, MASS is a vessel that can operate within a maritime environment using automated systems. This forms the foundation for the further regulatory and technological progress that aims to make the use of autonomous ships possible in the international maritime sphere.[2].

Even though many autonomous vessels have undergone prototype demonstrations, none have successfully achieved the IMO's complete safety and operational standards for full certification. The International Maritime Organization (IMO) has the role of developing guidelines and regulations concerning the safe and effective functioning of autonomous vessels. Even as late as 2024, autonomous vessels have yet to receive full approval from the international maritime organization for commercial utilization.

The implementation of robust autonomous ship systems is anticipated to significantly reduce the incidence of maritime accidents. This reduction is primarily attributed to the mitigation of human error, which is identified as the leading contributing factor in the majority of maritime incidents. As illustrated in Figure 1, the distribution of accidental events between 2011 and 2015 highlights the prevalence of human-related errors in naval accidents. Autonomous technologies are expected to enhance operational safety by minimizing reliance on human intervention and optimizing decision-making processes through advanced automation.

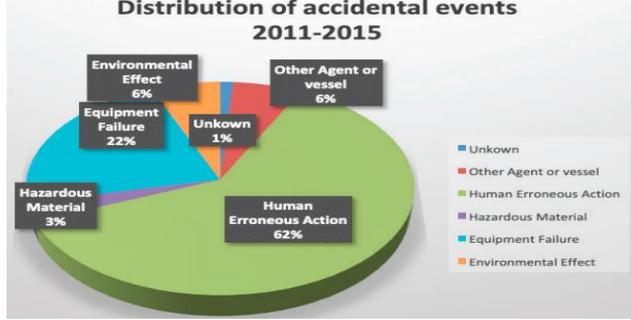

*Figure 1 Distribution of accidental events 2011-2015 (EMSA, 2016). adopted from [1].*

The high stakes of autonomous shipping are highlighted by the severe consequences of accidents involving large commercial vessels [3]. For instance, the 2021 grounding of the container ship Ever Given in the Suez Canal caused a six-day blockade, disrupting global logistics chains. The risks are further compounded by the massive size of some commercial ships, such as the Prelude FLNG, a floating liquefied natural gas facility measuring 488 meters in length, 74 meters in width, and with a gross tonnage of 600,000 tons.

In contrast to the relatively more forgiving environment of autonomous car testing, the maritime industry demands a more cautious and methodical approach. The current research aims to develop a methodology for guiding a small, rubber-hulled vessel navigating a dense obstacle environment, such as a naval port. The designed hardware and software will undergo extensive testing in real-world scenarios, spanning a significant number of nautical miles, to ensure the system's safety and robustness. This approach will enable the development of a reliable and efficient autonomous system that can be confidently applied to larger vessels.

## II. LITERATURE BACKGROUND

Autonomous shipping currently lacks comprehensive regulation by the International Association of Classification Societies (IACS) and the International Maritime Organization (IMO). As a result, the regulatory framework for such vessels is determined individually by flag states within their respective jurisdictions.

The IMO is actively developing a Maritime Autonomous Surface Ships (MASS) Code, with the non-mandatory version anticipated to be finalized and made available for voluntary implementation by 2025. Subsequently, an experience-building phase is planned between 2026 and 2028, aimed at refining the framework and informing the development of the Mandatory MASS Code. This mandatory version is expected to be adopted in 2028 and finalized by 2030, with its enforcement slated to commence in 2032[2] [4].

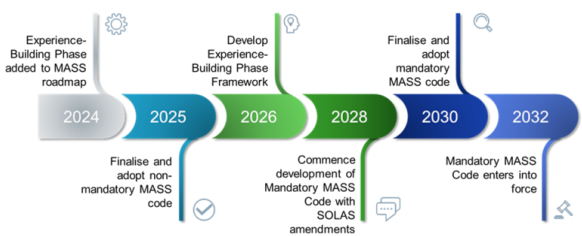

*Figure 2 MASS code timeline as suggested by IMO [4]*

There have been multiple attempts to create an autonomous ship like [5-7], Figure 3 shows a framework of a proposed method by [5].

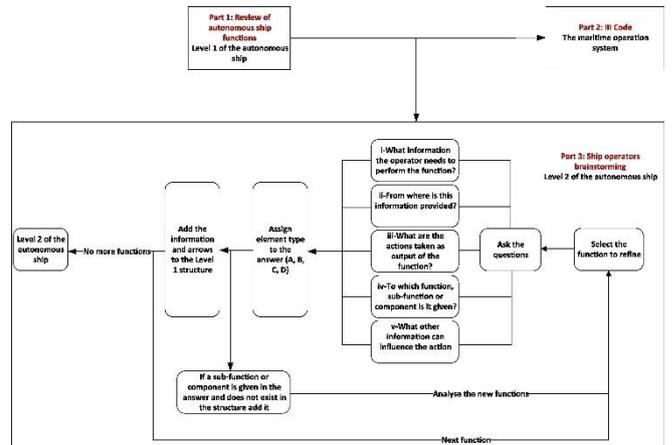

*Figure 3 framework to develop the hierarchical control structure of an autonomous ship adopted from [5].*

In general, there are 3 layers of autonomous control, the first layer is perception which is responsible for gathering information about the surrounding environment. The second layer cognition, which is responsible for processing the data and determining the action an autonomous vehicle should do. The third layer is the operation which is responsible for executing the command using actuators, Figure 4 is an illustration of this concept.

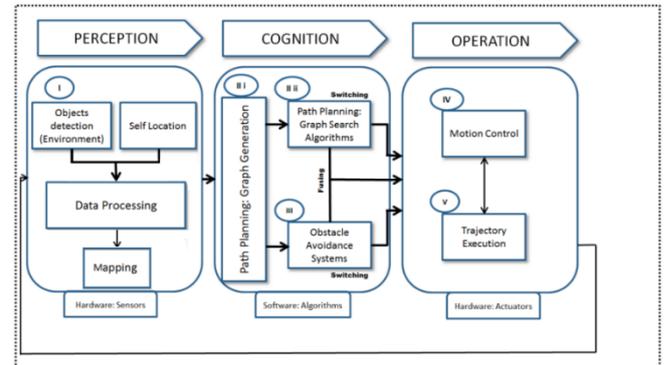

*Figure 4 Layers and phases of autonomous navigation. adopted from [8]*

## III. CONTROL SYSTEMS CRITERIA

The MASS control system must meet the following criteria to ensure safe navigation of the ship.

### A. Minimum update rate for navigation systems controller

Navigation systems should operate at an update rate optimized for the MASS characteristics to maintain effective situational awareness. The system must successfully execute the complete control loop: sensor data acquisition, data processing, decision-making, and actuator command transmission.

The minimum update rate can be derived from the following equation:

$$t_{sys.response} = 2 \times t_{sensor\ update} + t_{Mech.response} + t_{Mech.response} + \text{emergency stop}$$

Consider a representative case with the following parameters:

$v$ : Vessel speed = 10 m/s ($\approx$19.4 knots)

$d_{LIDAR}$ LIDAR range = 100 meters

$d_{emergency\ stop}$: Emergency stop distance = 10 meters

$t_{emergency\ stop}$: Emergency stop interval = 1s

$t_{sensor\ update}$ : Sensor update interval = 200ms (5 Hz)

$t_{Mech.\ response}$ : Mechanical response time = 0.5s

$t_{Eng.\ response}$ : Engine response time = 0.5s

At least two consecutive scans are needed for velocity vector determination, i.e.0.4s (more scans will yield more points, hence improving accuracy). In a worst-case head-on scenario with the other ship's speed of 10m/s, the relative velocity doubles to 20 m/s. At the sensor range of 100 meters, this yields 5 seconds until potential collision. Applying a safety factor of 2 reduces available reaction time to 2.5 seconds.

$$t_{sys.response} = 2 \times 0.2 + 0.5 + 0.5 + 1 = 2.4s$$

The remaining margin (0.1s) determines the minimum required update rate of 10Hz. This accounts for trajectory calculation accuracy, environmental factors, and system response variations.

Of course, this represents an oversimplified example, utilized to demonstrate the fundamental relationship between Decision time $t_d$ and system parameters in determining the minimum safe update rate for navigation systems controllers. In real-world applications, additional factors must be considered, including:

- Variable vessel dynamics
- Complex multi-vessel interactions
- Environmental conditions and their effect on sensor performance
- Non-linear mechanical response characteristics
- Processing overhead for more sophisticated collision avoidance algorithms

This basic model provides a framework for understanding the critical timing constraints in MASS control systems while acknowledging that actual implementations require a more comprehensive analysis based on specific operational requirements and conditions.

### B. Required control system capabilities

Control systems must handle complex navigation, obstacle detection, and real-time decision-making under diverse maritime conditions. The system must be able to process data from surrounding obstacles up to the highest number recorded in that port multiplied by a safety factor, at a rate less than or equal to the minimum update rate specified for navigation systems controllers in in section III.A

### C. Redundant processors with automatic failover

The system should have two processors: one controlling (the main) while the other is watching (the backup) and evaluating the situation. The main controller runs more sophisticated software for fuel economy and optimal equipment usage. If the backup processor detects that the main processor is not sending data at the required rate or that the orders to actuators may cause potential danger, the backup will disconnect power from the main and take control of the boat. The backup software is simpler than the main and will navigate the ship to the nearest safe pickup/fix point while sending a message to the control station indicating the change in control. It may reconnect power to the main processor and return control if the main system functions properly.

### D. Manual override capability

If the onboard crew or remote operators determine that the controller is not acting appropriately, they can remotely control the boat and disengage any onboard control. Human operators must maintain the ability to override autonomous systems in emergencies.

### E. Remote monitoring and control interface

The system must support remote access for monitoring and intervention, ensuring operational transparency and safety. A graphical user interface must display current status information, including:

- Surrounding environment
- Sensor data
- Equipment status
- Proposed navigation paths
- System performance metrics

The interface must provide the ability to override autonomous control and operate the boat remotely when necessary.

## IV. CYBERSECURITY REQUIREMENTS

Because autonomous ships are vulnerable to cyber threats[9], robust cybersecurity measures are critical. There are common cybersecurity measures known. And other measures suggested by the author.

### A. common cybersecurity measures

there are many cybersecurity measures applied. The measures differ by the importance of the system. In general, essential cybersecurity measures include:

1. Mandatory Security Protocols which are comprehensive sets of security measures must protect systems and data from unauthorized access.
2. All data transmissions, including navigation and operational instructions, must be encrypted to prevent interception using End-to-End Encryption.
3. Access to control and monitoring systems must require multi-factor authentication to enhance security.
4. Frequent evaluations must identify and address vulnerabilities in the system. The common method is by conducting regular security audits.
5. Systems must monitor for unauthorized access or suspicious activity in real time. This is known as automated intrusion detection.
6. The system must verify software integrity during boot and runtime to prevent unauthorized modifications. i.e. secure boot and runtime verification
7. Updates must be implemented promptly to address vulnerabilities and improve system functionality. This update should be regular.

### B. Additional Security Measures

Beyond standard cybersecurity practices, the following measures are essential from the point of view of the author:

*1) Disable Backup Controller Remote Upgrading*

As described in section III.C while the main controller's software may be upgraded remotely, the backup controller should only be upgradable through physical ports (serial, USB, or Ethernet). This measure prevents remote hackers from gaining full control of the vessel even if they breach all other security measures. The backup controller will detect deviations from the required parameters and disconnect the main controller, protecting the boat from unauthorized access.

Unauthorized access can take many forms. E.g. redirecting to another unwanted point, or modifying software to disable a safety feature to cause an accident like the famous case of the 2014 Jeep Cherokee a remote attacker could leverage a vulnerability in Uconnect to hack into a car's systems and perform various actions, from taking over the infotainment system to killing the engine and disabling the brake[10]…. etc.

At the moment the backup will notice the deviation from the required, the main will be disconnected, protecting the boat from any unauthorized access.

*2) Physically securing the backup controller.*

The backup controller should be physically separated from the main controller and secured behind robust physical barriers. This measure delays potential unauthorized physical access, providing time for vessel operators to respond to security threats.

## V. SUGGESTED SYSTEM LAYOUT

a suggested system layout that fulfills the requirement suggested by the author can be found in Figure 5

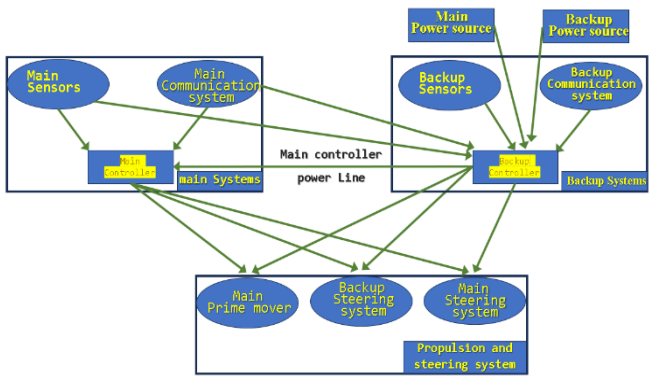

*Figure 5 Proposed MASS system layout.*

Maritime Autonomous Surface Ships (MASS) require comprehensive redundancy strategies to ensure operational reliability and safety. The proposed architecture implements a dual-controller system with hierarchical failover mechanisms.

The system employs two independent controllers:

- Primary Controller: Executes sophisticated navigation algorithms optimized for operational efficiency
- Secondary Controller: Maintains basic but robust navigation capabilities with simplified software architecture

Failover Mechanism

The failover system operates through the following hierarchy:

1. Normal Operation:
   - A primary controller manages all ship functions
   - A secondary controller monitors system parameters
   - The main communication link enables remote oversight
2. Level 1 main component Failure (Communication/steering/sensors...etc.):
   - Upon main system failure
   - The primary controller switches to the backup system.
   - The system maintains full operational capabilities.
3. Level 2 Failure (Primary Controller):
   - Triggered by:
     - Deviation from operational parameters
     - Irregular behavior detection
   - Secondary controller:
     - Physically disconnects primary controller power supply
     - Assumes control of all critical systems
     - Initiates safe-mode navigation protocols

This architecture ensures system resilience through:

- Physical isolation of redundant components
- Independent power supplies
- Separate communication channels
- Hierarchical control transfer protocols

## VI. SENSOR SYSTEMS AND PERCEPTION TECHNOLOGIES

### A. LiDAR (Light Detection and Ranging)

LiDAR sensors play a critical role in MASS perception systems by creating detailed 3D maps of the ship's surroundings. Their ability to provide accurate distance measurements and spatial awareness makes them particularly valuable in congested or complex maritime environments, such as ports or narrow waterways.

LiDAR can detect small objects that may not be visible to radar or cameras, enhancing the ship's ability to navigate safely. Additionally, these systems remain effective in low-visibility conditions, such as fog or darkness, where traditional visual navigation may be compromised. However, LiDAR performance can be affected by adverse weather conditions, such as heavy rain, fog, or snow, which may scatter or absorb the laser beams.

The LiDAR selection process must consider:
- Vessel speed requirements
- Surrounding vessel speeds
- Operational climate conditions
- Required detection range
- Weather resistance capabilities

The LiDAR should cover the 360° view field of the ship, in some cases, there should be a degree of overlap between sectors to ensure the discovery of all objects in the required range.

*B. Radar Systems*

Radar systems are essential for long-range detection in maritime environments, enabling autonomous ships to identify and track other vessels, landmasses, and large obstacles at distances of up to several nautical miles. Radar works by emitting radio waves and analyzing the reflected signals to determine the position, speed, and direction of objects in the ship's vicinity[11].

This capability is particularly important in open waters, where early detection of potential hazards is critical for safe navigation. However, radar systems may have limitations in detecting small or non-metallic objects, such as wooden boats or floating debris. Therefore, they should be integrated with LiDAR and visual sensors for comprehensive coverage.

*C. Cameras and Computer Vision*

High-resolution cameras, combined with advanced computer vision algorithms, form a cornerstone of MASS perception systems. These systems:
- Capture visual data from the ship's surroundings
- Process information using machine learning and AI techniques
- Identify and classify objects, including vessels, buoys, and navigational markers
- Enable real-time tracking of moving objects
- Support trajectory prediction and collision avoidance

Computer vision provides rich visual information, particularly useful for identifying small or irregularly shaped obstacles that may escape radar or LiDAR detection. Advanced algorithms can enhance navigation in low-visibility conditions through image enhancement techniques and thermal imaging. However, camera system performance may be affected by environmental factors such as glare, reflections, or water spray, necessitating integration with other sensor technologies.

*D. GPS and INS*

*1) System Overview*

Global Positioning System (GPS) technology provides precise location data for autonomous ships, enabling accurate determination of position, speed, and heading. Inertial Navigation Systems (INS) complement GPS by providing continuous position and orientation data, particularly valuable in environments where GPS signals may be compromised or unavailable.

*2) Cross-Verification Module*

Inertial Navigation Systems (INS) complement GPS by providing continuous position and orientation data, even in environments where GPS signals may be unavailable or unreliable, such as in tunnels, under bridges, or during satellite signal interference. Together, GPS and INS form a robust positioning system that is essential for the safe operation of autonomous ships. However, the integration of these systems requires careful calibration and error correction to ensure data accuracy and reliability.

From the author's perspective, a software module should be implemented to cross-check the position data provided by the Global Positioning System (GPS) and the Inertial Navigation System (INS). If both systems are available and their reported positions are close to each other, the GPS position should be prioritized due to its higher accuracy under normal conditions. However, if there is a significant discrepancy between the two positions, the system should evaluate which position is more likely to be correct based on the last registered position and the logical kinematic behavior of the ship (e.g., speed, heading, and acceleration). The position that aligns more closely with these parameters should be selected, while the other is discarded. This approach ensures that the ship's navigation system remains reliable even in the event of sensor errors or external interference.

Position discrepancies between GPS and INS can arise from various scenarios. The most common reason is that GPS signals are vulnerable to jamming, spoofing, or manipulation, which can lead to incorrect position data [12, 13]. Jamming occurs when malicious actors emit radio frequency noise to disrupt GPS signals, while spoofing involves broadcasting fake GPS signals to deceive the receiver causing it to be set to a wrong trajectory[14] [15].

An $80 million yacht, was taken control of in the Mediterranean Sea by a Texas University researchers' group by changing the trajectory of the yacht for about 100 m[16]. Such incidents highlight the need for robust cross-verification mechanisms between GPS and INS. In cases where GPS data is compromised, the INS can provide a reliable fallback, as it operates independently of external signals and relies on internal sensors (e.g., accelerometers and gyroscopes) to calculate the ship's position.

*3) INS System Malfunction*

While INS is highly reliable in the short term, it is prone to drift over time due to the accumulation of errors in its sensors. For instance, small inaccuracies in accelerometer or gyroscope measurements can compound over hours or days, leading to significant deviations in the calculated position. A study published in the Journal of Navigation (2021) found that INS drift rates can range from 0.1 to 1.0 nautical miles per hour, depending on the quality of the sensors and the calibration process.

In cases where the INS malfunctions or experiences excessive drift, the GPS can serve as a corrective reference. For example, if the INS reports a position that is inconsistent with the ship's last known location or kinematic behavior (e.g., an impossible speed or heading change), the system should prioritize the GPS data. Additionally, regular recalibration of the INS using GPS data can help mitigate drift and maintain accuracy over extended periods.

To address these challenges, the author proposes implementing a software module that continuously monitors and compares the position data from both GPS and INS. The module should include the following features:

Threshold-Based Validation: Define a threshold for acceptable differences between GPS and INS positions. If the difference exceeds this threshold, the system should flag the discrepancy and initiate a cross-verification process.

Kinematic Consistency Check: Evaluate the reported positions against the ship's last known location, speed, and heading to determine which data source is more reliable.

Error Logging and Alerts: Log discrepancies and trigger alerts for further investigation by the ship's operators or remote monitoring team.

Fallback Mechanisms: In cases where both systems report inconsistent data, the system should rely on alternative navigation methods, such as visual landmarks compass and speed logger, until the issue is resolved.

By incorporating these features, the proposed software module can enhance the reliability and resilience of autonomous ship navigation systems, ensuring safe and accurate operation even in challenging conditions.

*E. AIS (Automatic Identification System)*

The Automatic Identification System (AIS) is a critical communication tool for maritime navigation, enabling ships to exchange real-time information about their position, speed, course, and identity. AIS operates by transmitting and receiving VHF radio signals, which are then displayed on electronic chart systems or navigation displays. This allows autonomous ships to detect and track nearby vessels, predict their movements, and avoid potential collisions.

AIS is particularly valuable in congested or high-traffic areas, such as ports, harbors, and shipping lanes, where situational awareness is essential for safe navigation. By integrating AIS data with other sensor inputs, such as radar and LiDAR, autonomous ships can enhance their ability to make informed decisions and respond to dynamic maritime environments. However, AIS has limitations, such as reliance on other vessels to transmit accurate data and potential vulnerabilities to spoofing or cyberattacks. To address these challenges, AIS is often used in conjunction with other communication and sensor systems to ensure comprehensive situational awareness.

*F. Sensor Systems Integration (sensor fusion)*

The effective operation of Maritime Autonomous Surface Ships (MASS) relies on the integration of multiple sensor technologies, including LiDAR, radar, cameras, GPS, INS, and AIS. Each sensor system provides unique capabilities and data, which are fused using advanced algorithms to create a comprehensive perception of the ship's environment. This sensor fusion approach enhances the accuracy, reliability, and robustness of autonomous navigation systems, enabling ships to operate safely in a wide range of maritime conditions. In Figure 6 a schematic draw of MASS sensors layer.

For example, while LiDAR provides high-resolution 3D maps of the immediate surroundings, radar offers long-range detection of distant objects, and cameras deliver detailed visual information. GPS and INS ensure precise positioning, while AIS facilitates communication with other vessels. By combining these technologies, autonomous ships can achieve a level of situational awareness that surpasses traditional manned vessels, paving the way for safer and more efficient maritime operations.

Effective MASS operation requires sophisticated integration of multiple sensor technologies. This integration should:

- Combine data from all available sensors
- Apply advanced fusion algorithms
- Create comprehensive environmental awareness
- Enhance navigation accuracy
- Provide redundancy for critical systems
- Enable robust decision-making capabilities

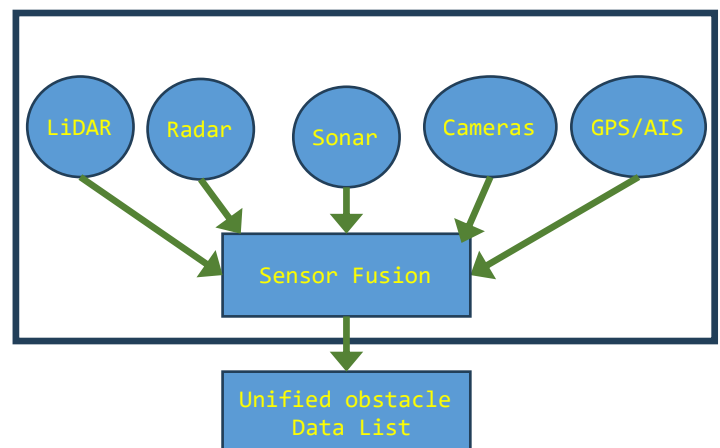

*Figure 6 MASS system sensor layer*

## VII. MASS COLLISION MODELS FOR COMPENSATION

The models for compensation in the event of a collision involving Maritime Autonomous Surface Ships (MASS)

remain a developing area of maritime law[17] [18] [19]. as traditional liability frameworks may not fully address the complexities introduced by autonomous systems. Generally, compensation mechanisms are rooted in international conventions, such as the Convention on Limitation of Liability for Maritime Claims (LLMC), the International Regulations for Preventing Collisions at Sea (COLREGs), and relevant national laws. The following are the common compensation models for MASS Collisions:

*A. Operator Liability Model*

In cases where an autonomous ship is operated remotely, the responsibility for collisions often falls on the operator or the company managing the vessel. Liability may be determined based on the operator's negligence or failure to properly monitor and control the ship[20].

*B. Manufacturer or Software Developer Liability:*

If the collision results from a failure in the ship's autonomous system, liability could shift to the manufacturer or developer of the software or hardware. This is akin to the principles of product liability in other industries, where defects or malfunctions lead to accidents.

*C. Strict Liability*

Some legal frameworks propose strict liability for MASS collisions, meaning the shipowner or operator would be held liable regardless of fault. This simplifies compensation but may impose significant financial risks on stakeholders.

*D. Insurance-Based Compensation*

Maritime insurers are adapting their policies to include coverage for autonomous operations. In this model, compensation would be handled through insurance claims, with premiums potentially adjusted to reflect the risks associated with autonomous systems.

*E. Shared or Proportional Liability*

In scenarios where both traditional ships and MASS are involved in a collision, liability could be distributed proportionally based on fault, as determined by investigations adhering to COLREGs and other maritime laws.

*F. Challenges and Considerations*

   *1) Allocation of Fault*

Establishing fault is complex for autonomous systems, especially if decisions are made by artificial intelligence without direct human input.

   *2) Regulatory Gaps*

Current international conventions may not explicitly address the unique aspects of MASS, requiring amendments or new legal instruments.

   *3) Experience-Building Phase*

As part of the IMO's ongoing efforts, the experience-building phase for MASS operations will provide insights into practical compensation models and inform future regulations.

## VIII. ECONOMIC AND ENVIRONMENTAL CONSIDERATIONS

Autonomous ships offer significant economic benefits, including reduced labor costs, optimized fuel consumption, and improved operational efficiency. However, their adoption may lead to job displacement and require new business models in the shipping industry.

From an environmental perspective, autonomous ships have the potential to reduce emissions through optimized routing and hybrid propulsion systems. However, the risks of accidents causing ecological damage must be addressed through rigorous testing and fail-safe mechanisms.

## IX. THE SUGGESTED FRAMEWORK

The certification process for autonomous ships should follow a multi-stage approach, beginning with trials within a country's territorial waters before seeking global IMO approval. This approach acknowledges that early phases may reveal software and hardware issues that could potentially cause problems, as demonstrated by incidents like the Suez Canal blockage. Each country should review the design and assess potential threats and benefits before approving subsequent trial stages.

The proposed certification process consists of six phases:

*A. Path plan algorithm selection/creation.*

The first step is to select a path plan algorithm from already created algorithms e.g. [21] [22] [23] [24] or design a path plan algorithm to be used onboard the autonomous ship. Then, a rubber boat model is created. Then a controller for this model is designed e.g. [25, 26] [27]. A simulation system must be created or selected to test the path plan and the ship guidance and control algorithms[29, 28].

The system should start from simple cases to real ports port traffic cases[30]. The testing will be as follows:

1- The ships entering/leaving the port as recorded by the Automatic Identification System IS system installed at the port are set as the moving obstacles in the system.

2- A start point and an endpoint are chosen on the port for the boat to create a valid path.

3- The path generated is checked by simulation if at any time the distance between the boat and any other is less than a certain value.

4- If a case is planned successfully, then the nested case is tested until all cases are tested successfully.

5- If one or more case test results were not safe by the simulation, then this case is marked and examined by the algorithm creator to fix it, or if there were too many errors select another algorithm.

6- The test cases should be all solved by the same version of the software, if one or more test cases failed and modifications were done to the system the test should be repeated. This is due to the author's experience, if a modification was done to fix a safety issue in a certain case, the modification may affect other parts of the program and cause other cases to fail even if the previous versions passed successfully.

- Stage Authorization:

    The marine authority of the state in which sea trials/operations are to be done.

- Stage Success criteria:

    Zero fail in the HIL simulation of the test cases obtained from real Port AIS with paths summation equal to or more than 5000 nautical miles.

- Human Role:

    Check manually some of the paths created to ensure good working of both the path plan and path check algorithms operation.

- Objectives:

    1. Evaluate the speed at which the system responds and operates.
    2. Evaluate the ability of the system to work in real-time.
    3. Collect data on system performance and environmental interactions.

- Steps:

    1. Creating a basic rubber boat model
    2. Developing or selecting a simulation system
    3. Testing path planning and ship guidance algorithms
    4. Progressing from simple cases to complex port traffic scenarios

- Testing Methodology:

    1. Use recorded AIS data from ports to create realistic moving obstacles
    2. Select start and end points for navigation testing
    3. Create a simulation system to verify path safety.
    4. Evaluate the distance between vessels, if below the set distance for each vessel, record safety violations
    5. Document and analyze any safety violations
    6. Modify algorithms as needed to address issues

*B. Ship control system design and testing by creating Hardware In Loop HIL system.*

Upon the completion of the path plan and ship guidance and control software, the algorithm should be implemented into a hardware controller. This controller should be tested by creating hardware in the loop system HIL[31] [32]. The created HIL will test the system's ability to work in real-time using the same steps mentioned earlier.

- Stage Authorization:

    The firm management.

- Stage Success criteria:

    Zero fail in the hardware in loop simulation of the test cases obtained from real Port AIS with paths summation equal to or more than 5000 nautical miles.

- Human Role:

    Check manually some of the paths created to ensure good working of both the path plan and path check algorithms operation.

- Objectives:

    1. Evaluate collision avoidance algorithms and emergency response protocols.
    2. Collect data on system performance and environmental interactions.

- Steps:

    1. Modify the software code developed in the previous step to work on a controller.
    2. Using the created simulation system create a hardware in the loop that receives actuators commands and uses them to create sensor data.
    3. Verify controller path safety through simulation
    4. Evaluate minimum distance requirements between vessels
    5. Document and analyze any safety violations
    6. Modify algorithms as needed to address issues

- Testing Methodology:

    1. Create Hardware in the loop to verify the controller's good operation.
    2. Use recorded AIS data from ports to create realistic moving obstacles
    3. Select start and end points for navigation testing
    4. Choose some test cases randomly and check the path manually to verify the good operation of the simulation and controller.
    5. Evaluate the distance between vessels and boats, if below the set distance for each vessel, record safety violations.
    6. Document and analyze any safety violations
    7. Modify controller software as needed to address issues

*C. Initial trials*

After the successful design and test of the controller using HIL, the design and prototyping of an autonomous rubber small boat is feasible. The boat can be used for delivery surveillance or environmental control outside the port, e.g., in the waiting area or port perimeter. There are two benefits. First, it will do a job, thus saving money. Second, doing a job requires continuous change which is important to test validity.

This phase should span approximately 15,000 nautical miles, with permission granted by a single country for a defined area. Any software or hardware bugs encountered should be addressed, and the trials repeated.

It is the author's point of view that the boat at the initial trial stage should not require other vessels nearby to do international rules to prevent collisions at sea maneuvers. The reason is at this stage the design robustness is not guaranteed or the other vessels may not recognize it. So, other ships nearby may be in the situation to do maneuvers that may not be safe for them and hence cause accidents. So, the system should act on it is own without requiring cooperation from surrounding/nearby ships which is even more difficult than the normal state. If the boat can achieve safety without the cooperation of the surrounding ships, then it can achieve safety with it, but not the opposite.

The ship design and working mechanisms may not be mature enough at this stage so, no human should not be onboard watching or ready to take control of the system.

- Stage Authorization

    The marine authority of the state in which sea trials/operations are to be done.

- Human Role

A human is remotely monitoring the system ready to interfere if anything wrong happens or the situation assesses that there is a high risk.

- Stage Success criteria

    Zero fail in the hardware in loop simulation of the test cases obtained from real Port AIS with paths summation equal to or more than 5000 nautical miles.

- Objectives

    1. Validate basic autonomous navigation capabilities.
    2. Assess sensor performance and data acquisition systems.

- Performance Metrics:

    1. Collision avoidance success rate.
    2. Navigation accuracy and responsiveness.
    3. Sensor reliability and data quality.
    4. System stability and resilience in various weather conditions.

- Steps:

    1. Design and manufacture a prototype for a small rubber boat.
    2. Use the controller created earlier, as the small rubber boat controller.
    3. Assign missions to boats and monitor the performance by humans to switch to manual control if needed.

- Testing Methodology:

    1. Record the exact position as obtained from sensors (GPS/INS).
    2. Record Evaluate distance between vessels and boat, if below-set distance for each vessel, record safety violations.
    3. Document and analyze any safety violations.
    4. Modify controller software/hardware as needed to address issues.

*D. Small craft trials*

Upon successful completion of the initial trials, the next stage should involve a small craft (up to 15 meters in length) with a crew onboard, ready to intervene in case of failure. The craft should complete 15,000 nautical miles, including at least 150 port entries and departures.

- Stage Authorization:

    The marine authority of the state in which sea trials/operations are to be done.

- Human Role:

    A human must be onboard, watching over the system ready to interfere if anything wrong happened or the situation assessment that there is a high risk.

- Stage Success criteria:

    - Zero collision incidents
    - ≤ 0.1% navigation error rate
    - ≤ 1% system failure rate
    - 99.9% availability of critical systems

- Objectives:

    - Test the system's ability to operate in real-world scenarios.
    - Evaluate human-machine interaction and crew response times.
    - Assess the system's ability to adapt to unforeseen circumstances.
    - Collect data on system performance in a wider range of operational conditions.

- Performance Metrics:

    - Successful completion of predefined routes and maneuvers.
    - Timely and appropriate responses to navigational challenges.
    - Crew workload and situational awareness.
    - System reliability and maintainability.

- Steps:

    1. Design and manufacture a prototype for a small ship.
    2. Modify the controller created earlier to be used as the small ship controller.
    3. Assign missions to ship and monitor the performance by humans onboard to switch to manual control if needed.

- Testing Methodology:

    1. Record data obtained from sensors for each trip to be used as a reference when needed.
    2. Record and analyze performed maneuvers for safety and also path optimization.
    3. Evaluate the distance between vessels and boats, if below the set distance for each vessel, record safety violations.
    4. Document and analyze any safety violations.
    5. Modify controller software/hardware as needed to address issues.

*E. Medium-sized ship trials.*

Following the small craft trials, the next stage should involve a medium-sized ship (15 to 100 meters in length) completing a trip from one port to another. The system minimum of 150 port operations. The crew is needed for both observing the guidance and control systems and also to fix any problem in the equipment that may lead to problems.

- Stage Authorization:

    The marine authority of the state in which sea trials/operations are to be done. If

- Human Role:

    - Be ready to stop the system and manually steer the ship if there is a situation in the system displayed response may not be safe enough.
    - Fix any system peripheral device faults and monitor sensor performance.

- Stage Success criteria:

    - Zero collision incidents
    - ≤ 0.01% navigation error rate
    - ≤ 0.1% system failure rate
    - 99.99% availability of critical systems

- Objectives:

    - Test the system's capabilities in long-distance operations.
    - Evaluate the system's ability to handle complex maritime traffic.
    - Assess the impact of environmental factors on system performance.

- Collect data on fuel efficiency, emissions, and operational costs.
- Performance Metrics:
    - On-time arrival and fuel efficiency.
    - Compliance with maritime regulations and safety standards.
    - System reliability and fault tolerance during extended operations.
    - Crew workload and stress levels.
- Steps:
    1. Design and manufacture a prototype for a mediumship or modify an existing one.
    2. Modify the controller to be used as the mediumship controller.
    3. Assign missions to ship and monitor the performance by humans onboard to switch to manual control if needed.
    4. Evaluate the performance from the point of view of fuel efficiency, emissions, and operational costs.
- Testing Methodology:
    1. Record data obtained from sensors for each trip to be used as a reference when needed.
    2. Record and analyze performed maneuvers for safety and also path optimization.
    3. Document and analyze any safety violations.
    5. Create a statistical analysis of fuel efficiency, emissions, and operational costs in various conditions.
    4. Modify controller software/hardware as needed to address issues.

*F. IMO certification.*

Based on the successful completion of the previous three stages, the autonomous ship manufacturer can apply to the IMO for a test certificate. This final stage would involve a more extensive 50,000 nautical mile trial, potentially across multiple countries, subject to their agreement.

- Stage Authorization

    International Maritime Organization IMO

- Human Role

    Minimal crew with monitoring and intervention capabilities

- Stage Success criteria
    - Zero collision incidents
    - ≤ 0.001% navigation error rate
    - ≤ 0.05% system failure rate
    - 99.999% availability of critical systems

- Objectives
    - Demonstrate compliance with IMO regulations and safety standards.
    - Assess the long-term reliability and maintainability of the system.
    - Evaluate the economic and environmental benefits of autonomous shipping.
    - Collect data on system performance and operational effectiveness in real-world conditions.
- Performance Metrics:
    - Safety record and incident rate.
    - Operational efficiency and cost-effectiveness.
    - Environmental impact and emissions reduction.
    - Public acceptance and societal impact.
- Steps:
    1. Apply to IMO for a MASS license with the already tested ship and its test log.
    2. IMO and Classification Societies (IACS) will review the design to make sure it meets the IMO regulations. If valid a temporary Test license will be issued for specific areas and routes.
    3. IMO will assign a representative(s) to attend in person some sea trials and monitor other progress online.
    4. Upon the successful completion of the sea trials, a final permanent license to operate worldwide.
- Testing Methodology
    1. Record data obtained from sensors for each trip to be used as a reference when needed.
    2. Record and analyze performed maneuvers for safety and also path optimization.
    3. Document and analyze any safety violations or any case the ship failed to comply with the Regulation of Convention on Limitation of Liability for Maritime Claims (LLMC), the International Regulations for Preventing Collisions at Sea (COLREGs), or the International Maritime Organization (IMO).
    4. Repeat the sea trials until neither safety violations found nor rules fail to comply is recorded.

*G. Framework overview*

A summary of the suggested certification process can be found in Figure 7. In this figure, certification phases are illustrated starting from the initial phase (path plan algorithm) until reaching the final phase (full commercial deployment). Also, the success criteria upon which certification moves from one phase to the next are shown.

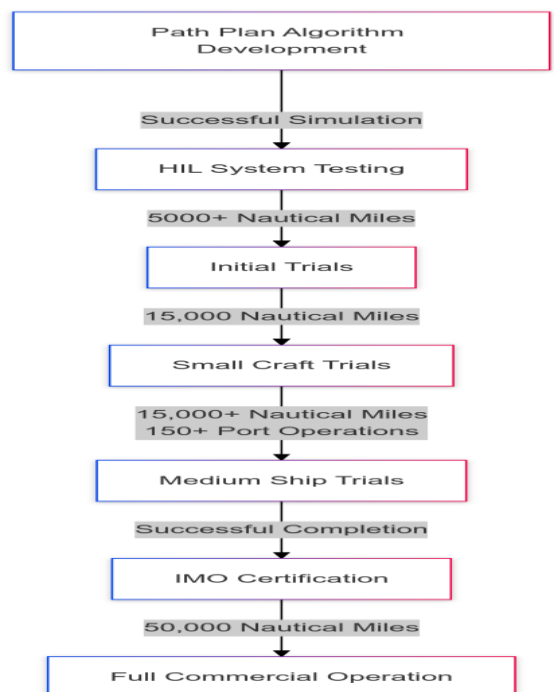

*Figure 7 proposed MASS Certification system flow chart.*

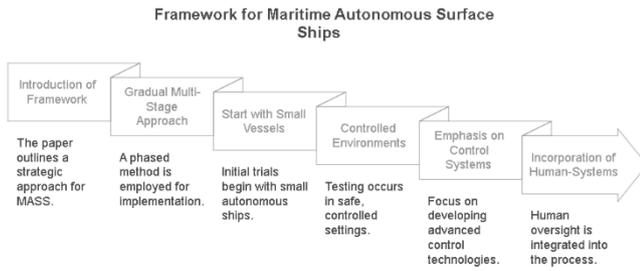

*Figure 8 proposed MASS Certification system key points.*

*1) Economic Viability and Scalable Implementation*

The proposed framework initiates autonomous vessel development with small-scale operations that serve dual purposes: system validation and revenue generation. This approach addresses both technical and financial sustainability aspects of autonomous vessel development.

*2) Initial Phase Revenue Generation*

The preliminary testing phase employs rubber-hulled vessels (< 5 meters) in commercially viable applications within controlled harbor environments. These operations can include:
  a. Environmental Monitoring Services
     o Water quality assessment
     o Marine pollution detection
     o Wildlife monitoring
     o Sediment sampling
  b. Port Logistics Support
     o Inter-terminal document delivery
     o Small cargo transport between anchored vessels
     o Supply delivery to vessels in waiting areas
     o Emergency equipment transport

This approach provides several advantages:
- Generates operational revenue during development
- Validates system reliability in real-world conditions
- Minimizes risk exposure due to vessel size
- Creates minimal interference with existing port operations
- Establishes stakeholder confidence through demonstrated performance

*3) Scalable Development Model*

The revenue generated from initial operations can partially offset development costs for subsequent phases, creating a self-sustaining development cycle:

Phase 1: Small vessel operations (< 5m)
  o Initial revenue generation
  o System validation
  o Risk minimization

Middle Phases: Medium vessel implementation (5-15m)
  o Expanded service capabilities
  o Increased revenue potential
  o Enhanced system complexity

Final Phase: Large vessel development
  o Overseas operations
  o Full autonomous capabilities
  o Commercial scale operations

This incremental approach provides several benefits:
- Reduced financial risk through staged investment
- Progressive validation of autonomous systems
- Continuous revenue generation throughout the development
- Stakeholder confidence building through demonstrated success
- Empirical data collection for regulatory compliance

The commercial viability of each phase supports the development of subsequent phases, creating a sustainable path toward full-scale autonomous shipping implementation. This approach aligns with both technical and financial risk management strategies while providing valuable operational data for system refinement and regulatory approval processes.

This multi-stage approach aims to ensure the safe and robust development of autonomous shipping technology before seeking global certification, minimizing the risks and potential disruptions to international maritime operations and at the same time allowing for some flexibility.

In all phases of trials and even after system deployment, there must be insurance on the ship/boat to pay for the damage it may cause in case of malfunctions.

## X. CONCLUSION

The integration of Maritime Autonomous Surface Ships (MASS) into global maritime operations is a complex and multifaceted challenge that requires a careful balance between technological innovation, regulatory oversight, and safety considerations. This paper has outlined a gradual, multi-stage certification process that begins with small-scale trials and progresses to full-scale international operations under the supervision of the International Maritime Organization (IMO). By addressing key issues such as control system reliability, cybersecurity, sensor integration, and redundancy mechanisms, the proposed framework aims to ensure the safe and efficient deployment of autonomous ships.

The economic and environmental benefits of MASS, including reduced labor costs, optimized fuel consumption, and lower emissions, are significant. However, these benefits must be weighed against potential risks, such as job displacement and the environmental impact of accidents. Furthermore, the evolving legal frameworks for liability and compensation in the event of collisions involving autonomous ships highlight the need for ongoing regulatory development and international cooperation.

As the maritime industry continues to navigate the uncharted waters of autonomous shipping, a cautious and methodical approach will be essential to minimize risks and ensure the successful integration of these advanced technologies into global trade. The proposed framework provides a roadmap for achieving this goal, while also allowing for flexibility and adaptation as new challenges and opportunities arise.

I. FUTURE WORK

While this paper has outlined a comprehensive framework for the certification and integration of Maritime Autonomous Surface Ships (MASS), several areas require further investigation. Future research should focus on advancing collision avoidance systems, enhancing cybersecurity measures, and developing robust redundancy mechanisms. Additionally, the integration of autonomous ships with port infrastructure, the environmental impact of autonomous shipping, and the socioeconomic implications of job displacement must be addressed. Collaborative efforts between industry stakeholders, regulatory bodies, and researchers are essential to ensure the safe and sustainable adoption of autonomous shipping technologies.

II. REFERENCES

1. Karlo Bratić, I.P., Srđan Vukša, Ladislav Stazić, *Review of Autonomous and Remotely Controlled Ships in Maritime Sector.* TRANSACTIONS ON MARITIME SCIENCE, 2019. **2**.
2. *Autonomous shipping*. 2025 [cited 2025; Available from: https://www.imo.org/en/MediaCentre/HotTopics/Pages/Autonomous-shipping.aspx?utm_source=chatgpt.com.
3. Mohamad Issa, A.I., Hussein Ibrahim and Patrick Rizk, *Maritime Autonomous Surface Ships Problems and Challenges Facing the Regulatory Process.* Sustainability, 2022. **14**.
4. *Autonomous and remotely-operated ships*. 2025 [cited 2025; Available from: https://www.dnv.com/maritime/autonomous-remotely-operated-ships/regulatory/?utm_source=chatgpt.com.
5. Meriam Chaala, Osiris A. Valdez Bandaa, Jon Arne Glomsrudb, Sunil Basneta, Spyros Hirdarisa,Pentti Kujalaa, *A framework to model the STPA hierarchical control structure of an autonomous ship.* Safety Science 2020. **132**.
6. Hans-Christoph BURMEISTER, W.B., Ørnulf Jan RØDSETH, Thomas PORATHE, *Autonomous Unmanned Merchant Vessel and its Contribution towards the e-Navigation Implementation: The MUNIN Perspective.* International Journal of e-Navigation and Maritime Economy, elsevier, 2014. **1**: p. 13.
7. Kim, J., *3D path planner of an autonomous underwater vehicle to track an emitter using frequency and azimuth–elevation angle measurements.* IET Radar, Sonar & Navigation, 2020. **14**(8): p. 8.
8. Marcus V. L. Carvalho, R.S., Leopoldo R. Yoshioka, *Autonomous Navigation and Collision Avoidance for Mobile Robots Classification and Review.* ResearchGate, 2024.
9. Rishikesh Sahaya, D.A.S.E., Weizhi Mengc,, Christian D. Jensenc,Michael Bruhn Barfod, *A comparative risk analysis on CyberShip system with STPA-Sec, STRIDE and CORAS.* Computers & Security, 2023. **128**.
10. Chris Valasek, C.M. *Remote Exploitation of an Unaltered Passenger Vehicle*. 2015; Available from: https://ioactive.com/pdfs/IOActive_Remote_Car_Hacking.pdf.
11. Tafsir Matin Johansson, D.D., Aspasia Pastra, *Maritime Robotics and Autonomous Systems Operations Exploring Pathways for Overcoming International Techno-Regulatory Data Barriers.* Journal Of Marine Science And Engineering, 2021. **9**.
12. Wenyi Wang, J.W., *GNSS induced spoofing simulation based on path planning.* IET Radar, Sonar & Navigation, 2021. **10**(1): p. 10.
13. S. Liu, X.C., H. Yang, Y. Shu, X. Weng, P. Guo, K.C. Zeng, G. Wang, Y. Yang. *Stars Can Tell: A Robust Method to Defend against GPS Spoofing Attacks using Off-the-shelf Chipset*. in *THE 30TH USENIX SECURITY SYMPOSIUM*. 2021.
14. Whelan J., A.A., Braverman J., El-Khatib K., *Threat analysis of a long range autonomous unmanned aerial system*, in *2020 International Conference on Computing and Information Technology*. 2020.
15. Daniel P. Shepard, J.A.B., and Todd E. Humphreys, Aaron A. Fansler. *Evaluation of Smart Grid and Civilian UAV Vulnerability to GPS Spoofing Attacks*. in *25th International Technical Meeting of the Satellite Division of the Institute of Navigation*. 2012.
16. Golijan, R. *Cheap GPS trick sends $80 million superyacht off course*. 2013; Available from: https://www.nbcnews.com/technolog/cheap-gps-trick-sends-80-million-superyacht-course-6c10796390.
17. Henrik Ringbom, E.R., Trond Solvang, *Autonomous Ships and the Law* IMLI Studies in International Maritime Law, ed. D.J. Attard. 2021, New York: Routledge.
18. *OUTCOME OF THE REGULATORY SCOPING EXERCISE FOR THE USE OF MARITIME AUTONOMOUS SURFACE SHIPS (MASS)*, IMO, Editor. 2021.
19. Petrig, A., *Maritime Security in the Age of Autonomous Ships.* SSRN 2023.
20. Ringbom, H., *Regulating Autonomous Ships—Concepts, Challenges and Precedents.* Ocean Development & International Law, 2019.
21. Alessandro Gasparetto, P.B., Albano Lanzutti, Renato Vidoni *Path Planning and Trajectory Planning Algorithms: a General Overview*. Motion and Operation Planning of Robotic Systems. Mechanisms and Machine Science. 2015, Switzerland: Springer.
22. Arnaoot, H.M., *Hexagon path planning algorithm.* IETRadar,Sonar&Navigation, 2022. **1**(17).
23. Hany Arnaoot, H.A.A., *Enhancing the Hexagon Path Planning Algorithm for Dense Obstacle*

*Environments (Iterative Hexagon Algorithm)* The International Journal of Telecommunications  IJT, 2024. **4**(2): p. 17.
24. M.S.Ganeshmurthy, D.G.R.S., *Path Planning Algorithm for Autonomous Mobile Robot in Dynamic Environment*, in *3rd International Conference on Signal Processing, Communication and Networking (ICSCN)* 2015 IEEE: Chennai, India. p. pp. 1-6.
25. TOMERA, M., *NONLINEAR CONTROLLER DESIGN OF A SHIP AUTOPILOT.* Int. J. Appl. Math. Comput. Sci., , 2010. **20**(2): p. 271–280.
26. Anish Pandey, D.R.P., *MATLAB Simulation for Mobile Robot Navigation with Hurdles in Cluttered Environment Using Minimum Rule Based Fuzzy Logic Controller.* 2nd International Conference on Innovations in Automation and Mechatronics Engineering,ICIAME 2014 , elsevier, 2014. **14**(-): p. 7.
27. Blomhoff, J.S., *Controller design for an unmanned surface vessel*, in *Department of Engineering Cybernetics*. 2007, Norwegian University of Science and Technology.
28. M. Mosleh, G.A.Z., Hany M. Arnaoot, *An Alternative Method to Simulate Three-Dimensional Point Scan Sensors Aboard Moving Vehicle* in *International Conference of Electrical Engineering*. 2018, Military Technical College, Ministry of defense: Cairo, Egypt.
29. Damitha Sandaruwan, N.K., Chamath Keppitiyagama and Rexy Rosa, *A Six Degrees of Freedom Ship Simulation System for Maritime Education*  The International Journal on Advances in ICT for Emerging Regions, 2010. **2**: p. 14.
30. Arnaoot, H.M., *An Approach for Real-Time Trajectory Planning in Dense obstacle Dynamic Environments* Academia Engineering, 2025. **1**(1).
31. Hany M. Arnaoot, G.A.Z., *Design of Hardware in the Loop System for Autonomous Ship Carrying Point Scan Sensor (Laser Measurement System).* AUI Research Journal, 2020. **1**: p. 26.
32. Khalid bin Hasnan, L.B.S.a.T.H. and *A Hardware-In-the-Loop Simulation and Test for Unmanned Ground Vehicle on Indoor Environment* 2012 International Workshop on Information and Electronics Engineering (IWIEE) 2012. **29**: p. 5.